\documentclass[fleqn,11pt,twoside]{article}

\usepackage{amsthm,amsthm,amssymb, color, xcolor,epsfig, graphics, subfigure}

\usepackage{amsmath, graphicx, latexsym, lscape, hyperref }
\usepackage{float}

\makeatletter
\newcommand{\copyrightnote}[2]{{\renewcommand{\thefootnote}{}
 \footnotetext{\small\it
\begin{flushleft}
 \copyright \ #1   #2  
\end{flushleft}}}}

\newcommand{\Name}[1]{\begin{flushleft}
                       \LARGE \bf #1
                       \end{flushleft}\vspace{-3mm}}

\newcommand{\Author}[1]{\begin{flushleft}
                       \it #1 \end{flushleft}}

\newcommand{\Address}[1]{\begin{flushleft}
                       \it #1 \end{flushleft}}

\newcommand{\Date}[1]{\begin{flushleft}
                      \small  \it #1 \end{flushleft}}

\newcommand{\evenhead}{Author \ name}
\newcommand{\oddhead}{Article \ name}

\renewcommand{\@evenhead}{
\hspace*{-3pt}\raisebox{-15pt}[\headheight][0pt]{\vbox{\hbox to \textwidth
{\thepage \hfil \evenhead}\vskip4pt \hrule}}}
\renewcommand{\@oddhead}{
\hspace*{-3pt}\raisebox{-15pt}[\headheight][0pt]{\vbox{\hbox to \textwidth
{\oddhead \hfil \thepage}\vskip4pt\hrule}}}
\renewcommand{\@evenfoot}{}
\renewcommand{\@oddfoot}{}

\long\def\@makecaption#1#2{%
  \vskip\abovecaptionskip
  \sbox\@tempboxa{\small \textbf{#1.}\ \ #2}%
  \ifdim \wd\@tempboxa >\hsize
    {\small \textbf{#1.}\ \ #2}\par
  \else
    \global \@minipagefalse
    \hb@xt@\hsize{\hfil\box\@tempboxa\hfil}%
  \fi
  \vskip\belowcaptionskip}

\newcommand{\JNMPnumberwithin}[3][\arabic]{%
  \@ifundefined{c@#2}{\@nocounterr{#2}}{%
    \@ifundefined{c@#3}{\@nocnterr{#3}}{%
      \@addtoreset{#2}{#3}%
      \@xp\xdef\csname the#2\endcsname{%
        \@xp\@nx\csname the#3\endcsname .\@nx#1{#2}}}}%
}

\newcommand{\resetfootnoterule} {
  \renewcommand\footnoterule{%
  \kern-3\p@
  \hrule\@width.4\columnwidth
  \kern2.6\p@}
}

\renewcommand{\footnoterule}{}

\makeatother

\theoremstyle{definition}

\begin{document}

\renewcommand{\evenhead}{ {\LARGE\textcolor{blue!10!black!40!green}{{\sf \ \ \ ]ocnmp[}}}\strut\hfill  
M Scalia, O Ragnisco, B Tirozzi and F Zullo
}
\renewcommand{\oddhead}{ {\LARGE\textcolor{blue!10!black!40!green}{{\sf ]ocnmp[}}}\ \ \ \ \  
The Volterra Integrable case}
%%%% Matter for the first page 
\thispagestyle{empty}
\newcommand{\FistPageHead}[3]{
\begin{flushleft}
\raisebox{8mm}[0pt][0pt]
{\footnotesize \sf
\parbox{150mm}{{Open Communications in Nonlinear Mathematical Physics}\ \  {\LARGE\textcolor{blue!10!black!40!green}{]ocnmp[}}
\ Vol.4 (2024) pp 
#2\hfill {\sc #3}}}\vspace{-13mm}
\end{flushleft}}

\FistPageHead{1}{\pageref{firstpage}--\pageref{lastpage}}{ \ \ Article}

\strut\hfill

\strut\hfill

\copyrightnote{The author(s). Distributed under a Creative Commons Attribution 4.0 International License}

\Name{The Volterra Integrable case. Novel analytical and numerical results.}

\Author{M. Scalia$^a$, O. Ragnisco$^b$, B. Tirozzi$^c$, F. Zullo$^d$}

\Address{$^a$Dipartimento di Matematica, Universit\`a degli Studi ``La Sapienza", P.le A. Moro 2, I-00185 Roma\\
\noindent    
$^b$Dipartimento di Matematica e Fisica, Universit\`a degli Studi ``Roma Tre" (retired), Via della Vasca Navale 84,00146 Roma\\
\noindent
$^c$ Dipartimento di Fisica, Universit\`a degli Studi ``La Sapienza", P.le A. Moro 2, I-00185 Roma\\
$^d$
DICATAM, Universit\`a degli Studi di Brescia, via Branze, 38 - 25123 Brescia, Italy \&\\
INFN, Milano Bicocca, Piazza della Scienza 3, Milano, 20126, Italy}
\Date{Received July 17, 2024; Accepted October 3, 2024}

\setcounter{equation}{0}

\begin{abstract}
\noindent 
In the present paper we reconsider the integrable case  of the Hamiltonian $N$-species Volterra system, as it has been introduced by Vito Volterra in 1937 and significantly enrich the results already published in the ArXiv in 2019 by two of the present authors (M. Scalia and O. Ragnisco).
In fact, we 
present a new approach to the construction of conserved quantities and comment about the solutions of the equations of motion; we display mostly new analytical and numerical results, starting from the classical predator-prey model and arriving at the general $N$-species model. 
\end{abstract}

\label{firstpage}

%%%% The Article text starts here

\section{Preface}
This paper is dedicated to the memory of one of the authors, namely our dearest colleague and friend Massimo Scalia, who sadly died  on December 10, 2023, after a dramatic car accident occurred at  a crossroad   not far from the centre of Rome. In the previous summer, Massimo had been deeply affected by a major personal tragedy, the death of his beloved companion (Adele Vannini)  who passed away after a long illness. He was able to react to this tremendous shock, fully devoting himself  to the two fields of interest that marked his whole life, politics and science. We could even say that they were not two different interests, inasmuch as, all along his career, he tirelessly struggled to establish a bridge between them.
The idea of resuming the brilliant approach to Mathematical Biology introduced by Vito Volterra  \cite{Volterra2}, recently revived by Giorgio Israel in his beautiful monographs \cite{Israel}, has in fact characterized several of  his recent papers, among which we just quote the most recent ones \cite{ECEC, Scalia 1, Scalia 2}. From \cite{Volterra2}, Massimo took the key idea that conflicting variables,  one economical and the other ecological,  could be thought  as forming a pair of conjugated variables (in the Hamiltonian language) in the spirit of the predator-prey model \cite{Volterra1}, following the ideas introduced by Goodwin already fifty years ago \cite{Goodwin}, in his class struggle model.
%applied to build the Goodwin model, but with a pair of variables
%only economic to describe an economic cycle \cite{Goodwin}, the so called
%, where a variable linked to the wage rate
%assumes the role of predator and the one giving employment rate
%is the prey.

The analogy could indeed be pushed forward to involve a larger number of competing species, so  deriving an   $N$-species generalization of the simplest two-species model. Once recast in a Hamiltonian form, it was then natural to ask whether also such an  enlarged model could enjoy the property of complete integrability. In fact we discovered that Volterra himself had already addressed  and solved this question, that amounted to require a specific (but simple) structure for the interaction matrix. The present article focusses on the study of the $N$-species Volterra model in the integrable case, and as such is a reformulation and  a significant extension of the preprints appeared  in the ArXiv
\cite{RS}, whose contents are here largely incorporated.

\section{Introduction.}
%The current twofold crisis, at a global level, of Economics and
%Environment has been deeply investigated and reported by many
%authors, some of which have deeply criticized the deafness of
%Economists with respect to the environmental crisis, that at most
%has been assessed for its dramatic consequences on GDP (see the
%well known ``Stern's Report" \cite{Stern}). The double crisis and its
%entanglement would request models, even before than a theory,
%able to put together economic and environmental variables in
%order to build a global stationary state to rule present predicament
%in the perspective of a sustainable scenario. The latter theme is not
%a news, several attempts having been realized starting from the
%seventies towards a definition of ``steady state"(\cite{Georgescu 1,Georgescu 2, Daly}) but it could be
%useful to face the problem with other scientific tools, as it has been
%recently proposed in \cite{Scalia1,Scalia 2,Scalia 3}, where the leading idea is to put together
%pairs, each constituted by an economic variable and an
%environmental one, that present a behavior of ``predator-prey" type,
%as it is suggested by some of the most important pairs one can
%select for the model. 
 
%%%%%%%%%%%%%%%%%%%%%%%%%%
%\sect{The Model}

%Here, aiming at giving a self-contained description  of our approach, we resume what two of the authors wrote in \cite{RS}. 
%\noindent
As well known, the original idea by Vito Volterra \cite{Volterra1} was that of
determining the evolution of a two species biological system, the
so-called predator-prey model, answering a question raised by
his  son in law, the biologist Umberto
d'Ancona \cite{dancona}, who was wondering why the total catch of selachians
(mostly sharks) was considerably raising during World War 1,
with respect to other more desirable kind of fishes, in
correspondence with the decrease of fishing activity \cite{Braun}. To answer
that question, Vito Volterra constructed a dynamical system that
enabled him to identify the essential features of what was going on,
elucidating the properties entailing the existence of a stable
equilibrium configuration and of periodic orbits, and unveiling the asymptotic behavior of the
system under general initial conditions. He quickly realized that
the predator-prey model was just the simplest example in a large
class of biological, or rather  ecological systems with pairwise
interaction. He was soon interested in understanding the
mathematical properties of the $N$ species pairwise interacting
model, and expend a considerable effort to find suitable
Lagrangian and Hamiltonian formulations, with the final aim of
achieving a description where the deep analogy with the well
established theory of mechanical systems stemming from the
Maupertuis  minimal action principle be made transparent. We
would say that not the whole Biological-Mechanical dictionary
that he proposed in his famous paper (dating back to 1937),
{\it Principes de Biologie Math\'ematique} \cite{Volterra2}, resisted the future
developments of both disciplines, and some of the notions he tried
to introduce look nowadays a bit artificial. But we believe that the
core of his  derivation is still alive, as it has been witnessed by 
very widespread applications over about a century in many
scientific research subjects, such as Demography, Biophysics, Biomedicine,
Ecology, Economics but also chemical reactions modeling and the theory of oriented directed
graphs. We notice that in \cite{Volterra2} his main aim was the formulation of this generalized model in  a Hamiltonian language, with the final purpose of elucidating  the algebraic conditions leading to a completely integrable model. Actually, the direction he chose, aiming to establish what he called the ``Three fundamental laws of biological fluctuations" \cite{Volterra2}, pp. 20-21, namely that of looking for  a conservative model, is certainly not the only possible generalization of the original predator-prey system. The current literature is quite rich of papers dealing with dissipative models, see for instance \cite{Telmo Jorge Lucas Peixe, LO, Baigent}.
 
Although \cite{Volterra2} has been inexplicably neglected by most of the researchers who worked on Lotka-Volterra systems in their various formulations, we consider it a seminal paper and take  it as our starting point. Accordingly, first of  we recall his approach and his  main results, and then propose  a novel approach to integrability, in order to extract new features of the model, first focussing on the three-species case and then generalizing it to an arbitrary number of species. To perform such task we stayed with the basic Volterra's  assumption that the interaction matrix $A$, see (\ref{model}), be skew-symmetric. We are aware that this assumption could sound unrealistic, but on the other hand, to the best of our knowledge, this property, or better its generalization, the so-called skew-symmetrizability, defined for instance in  \cite{Telmo Jorge Lucas Peixe} and in \cite {Baigent}, looks fundamental for dealing with conservative systems, and a fortiori with completely  integrable ones. 
\noindent
Our paper is organized as follows.
\noindent
In Section 2 we recall how Volterra constructed his Lagrangian and Hamiltonian formulation for the general $N$-species model, and we briefly comment on different Hamiltonian structures existing in the literature. Also, we present the Volterra's approach to complete integrability, and, as a by-product, we exhibit his Hamiltonian description of the predator-prey model; different, possibly more familiar, Hamiltonian  formulations, are also recalled here and more explicitly in the Appendix.  In the end, we discuss equilibrium configuration for $N$ larger than 2, starting with $N=3$, so introducing the content of the next section. 
\noindent 
 In Section 3 we present an analytical argument and a geometric evidence  to infer the compactness of the space of trajectories for $N=3$, and display a number of periodic examples. Further, a general argument to deal with the $N$-species problem is given, and a set of  independent integrals of motion is displayed. The proof of their involutivity is confined to Appendix A.
\noindent
Section 4  is the concluding one: we make  comments on  Volterra's results compared to ours and outline some possible future developments.
\section{The $N$-species system.}
 \noindent
The equations for the $N$-species Volterra System read
\begin{equation}
\dot{N}_r = \epsilon_r N_r + \sum_{s\ne r=1}^N A_{rs}N_rN_s ~~(r=1,\cdots,N)\label{model}
\end{equation}
In (\ref{model}) a dot on a function represents the time derivative and we have set all the parameters  introduced in \cite{Volterra1} $\beta_r =1 ~\forall r$ (which is not totally harmless from the biological point of view, as Volterra explains in the second paragraph  of the first part of his essay); $\epsilon_r$ are the natural growth coefficients of each species and $A_{rs} $ are interaction coefficients between species $r$ and species $s$ that account for the effects 
 of encountering between two individuals (more precisely, according to \cite{Volterra2}, $(1/\beta_r)A_{rs}N_rN_s$ denotes the decreasing  in unit time of the individuals of the specie $r$, while $(1/\beta_s)A_{sr}N_rN_s$ denotes the corresponding increasing of the species $s$).
 If  the matrix $A$, whose elements are $A_{rs}$, is nonsingular,
  then the system of equations defining  equilibrium configurations (other than the trivial one $\{N_r\} =\{0\}$), namely
 \begin{equation}
0= \epsilon_r + \sum_{s=1}^N A_{rs}N_s^{(0)} \label{equi}
\end{equation}
has a unique solution, say $N_r^{(0)}$, $r=1,\cdots,N$.  If, in addition, according to \cite{Volterra1}, we require $A$ to be skew-symmetric and  $N$  to be even, then the eigenvalues of $A$ will be purely imaginary and complex conjugate in pairs.  On the contrary, in the case of an odd number of species, a skewsymmetric $A$ will be singular and the system (\ref{equi}) will not have a single equilibrium solution, but possibly infinitely many.  What  is more important, however, in the biological context, is that  the roots of (\ref{equi}) be all positive. As a necessary condition,  the natural growth coefficients $\epsilon_r$ cannot have all the same sign. We emphasize that we will assume $N_r>0$  and  $\epsilon_r \ne 0~\forall r$ throughout the whole paper
 \subsection{Lagrangian and Hamiltonian formulation}
As many other researchers of his time, Volterra was feeling more assured if a phenomenon quantified by Mathematics could find an analogue with Mechanics, that moreover allowed resorting to the powerful formalism and theorems of the latter. To achieve this goal, Volterra introduced the quantity  of life for each species $r$ being defined as $q_r = \int_0^t N_r(\tau)d\tau$. The quantities of life were instrumental  for  the introduction of a biological, or rather   ecological, Lagrangian $\Phi$, defined as:
\begin{equation}
\Phi =  \sum_r \epsilon_r q_r +\sum_r  \dot{q}_r \log  \dot{q}_r - \frac{1}{2}\sum_{rs}A_{rs}\dot{q}_r q_s  \label{lagr1}
\end{equation}
In terms of (\ref{lagr1}), (\ref{model}) can be written as Euler-Lagrange equations
\begin{equation}
\frac{d}{dt}\frac{\partial \Phi }{\partial \dot{q}_r} - \frac{\partial \Phi}{\partial q_r} = 0 \label{EL}
\end{equation}
\noindent
yielding the ODEs
\begin{equation}
\ddot{q}_r=(\epsilon_r +\sum_s A_{sr}\dot{q}_s)\dot{q}_r\label{ODEs}
\end{equation}
\noindent
which are just (\ref{model}), up to the substitution $N_r=\dot{q}_r$. %the $\prime$ %denoting time-derivative. 
%In the following we will change the notations, by setting:
%\begin{equation}
%X_r := q_r; ~~\dot{X}_r := \dot q_r\label{Lagr}
%\end{equation}
%In other words, we turn to the standard notation of Lagrangian Mechanics. 
The transition from the Lagrangian to the Hamiltonian description is performed by Volterra in the usual way.
%\subsect{From Lagrange to Hamilton}
\noindent
The linear momenta, canonically conjugated to the quantities of life, are defined as
\begin{equation} 
p_r =\frac{\partial \Phi}{\partial  \dot q_r}= \log \dot q_r  +1 - \frac{1}{2}\sum_s A_{rs}q_s\label{momenta}
\end{equation}
\noindent
whence
\begin{equation}
\dot q_r = \exp(p_r -1+\frac{1}{2}\sum_s A_{rs}q_s)
\end{equation}
\noindent
Through a transformation of Legendre type Volterra defines the Hamiltonian 
\begin{equation}
{\mathcal H} = \Phi - \sum_r \dot q_r  p_r\label{Ham0}
\end{equation}
A straightforward calculation allows to rewrite (\ref{Ham0}) in the form:
\begin{equation}
{\mathcal H} = \sum_r\epsilon_rq_r-\dot q_r =\sum_r\epsilon_r q_r 
- \exp(p_r -1+\frac{1}{2}\sum_s A_{rs}q_s)\label{Ham}
\end{equation}
\noindent
Note that, in terms of the original ecological variables the expression $\sum_r\epsilon_rq_r-\dot q_r $
takes the form:
\begin{equation}
\sum_r\epsilon_rq_r-\dot q_r = \sum_r\epsilon_r \int_0^t dt'N_r(t')- N_r \label{Hamecol}
\end{equation}
\noindent
Volterra \cite{Volterra2} showed that (\ref{model}) can be written in the standard Hamiltonian form 
\begin{equation}
\dot q_r = - \frac{\partial {\mathcal H} }{\partial p_r}\label{Ham1}
\end{equation}
\begin{equation}
\dot p_r =\frac{\partial {\mathcal H} }{\partial q_r}\label{Ham2}
\end{equation}
\noindent
It is easily seen (see again [4]) that the  Hamiltonian system (\ref{Ham})-(\ref{Ham2}) has the following $N$ independent non autonomous integrals of motion:
\begin{equation}
{\mathcal H_r} = \frac{p_r - \frac{1}{2}\sum_s A_{rs}q_s}{\epsilon_r} -t ~~r=1,\cdots,N. \label{H_r}
\end{equation}
\noindent
whence  one can select $N-1$ autonomous integrals by taking for instance ${\mathcal H_{1,r}}\equiv {\mathcal H_r}-{\mathcal H_1}$, and have a complete set by adding the Volterra $N-1$ species Hamiltonian (\ref{Ham}). \\
We notice that the choice of the signs in (\ref{Ham1}) and (\ref{Ham2}) is opposite with respect to the standard approach found in literature (see e.g. \cite{Arnold, Marsden, goldstein}). Also, it would be preferable to have a minus sign in the definition of the Hamilton function $\mathcal{H}$ (\ref{Hamecol}), so that $\mathcal{H}(t=0)=\sum_r N_r$. We however prefer here to keep the notation used by Volterra himself in \cite{Volterra2}. Also, the term $+1$ appearing in (\ref{momenta}) is not crucial and could be replaced by any constant factor for example by adding in the Lagrangian (\ref{lagr1}) a combination of the variables $\dot{q}_r$.
\noindent
A more modern approach to the Hamiltonian structure underlying the generalized Volterra system can be  found for instance in  \cite{Telmo Jorge Lucas Peixe, LO, Baigent} where a Poisson morphism is established between  the original system, living in ${\mathbb R}^N$ and equipped with a quadratic Poisson structure, and  then one recast, after Volterra, in a canonical Hamiltonian form  and thus living in   ${\mathbb R}^{2N}$, see also \cite{Plank}.
We will come back to such Hamiltonian formulations, possibly more widely used than Volterra's
one, in subsection (2.3) and in the Appendix. Let us remark that different Hamiltonian descriptions of the Lotka-Volterra exist (see e.g. \cite{LO}, where a map between them is also given): if the Hamiltonian (\ref{Ham}) is rewritten in terms of the numerosities $N_r$ with respect to the Poisson bracket given in \cite{LO}, then the standard logarithmic terms of the numerosities of the populations appears.
\noindent
In our opinion, the question whether there exists a special form of the matrix elements $A_{rs}$  entailing involutivity of the complete set of integrals of motion $({\mathcal H}, {\mathcal H_{1,r}})$ is a  relevant one to ask both mathematically and from the point of view of applications: indeed, broadly speaking,  if on one hand the integrability structure leads to a very rich and assorted type of dynamical behavior, on the other hand it is expected that these properties may give a useful insight to the understanding of a wide variety of phenomena in a number of different fields. It turns out that this form has been found by Volterra himself \cite{Volterra2} and is the following:
\begin{equation}
A_{rs} =\epsilon_r \epsilon_s (B_r-B_s)~~~r,s=1,\cdots N\label{matrix}	
\end{equation}			
\noindent
where $N$ is the number of competing populations and the $B_r$ are  distinct real  numbers. Clearly, (\ref{matrix}) can  be cast in the compact form:
\begin{equation}
A = \{A_{rs}\} = [B,\epsilon \otimes \epsilon] \label{A}
\end{equation}
where $B$ $=diag(B_1,\cdots,B_N)$, and $\epsilon$ is the vector $(\epsilon_1,\cdots,\epsilon_N)^t$, meaning that $ A$ is the commutator of a diagonal matrix with distinct entries and a rank one matrix.
\medskip 
 \subsection{A note about the equilibrium conditions.}\label{sub3.3}
 \noindent
So, if we require integrability,  (\ref{A}) shows that the invertibility of $A$ has to be given up  for $N>2$. Indeed, in the integrable case Ker$(A)$ has dimension $N-2$, and correspondingly its range is two-dimensional. Accordingly,  the equilibrium conditions (\ref{equi}) read:
$$\sum_{s=1}^N\epsilon_s(B_s-B_r)N^{0}_s =1$$
\noindent
 where we have denoted by $N^{0}_s (s=1,\cdots,N)$ the equilibrium population numerosities, implying that the equilibrium configuration be defined as the intersection of the two hyperplanes 
\noindent
\begin{eqnarray}                
\sum_s\epsilon_sN^{0}_s=0\\	
\sum_s \epsilon_s B_sN^{0}_s=1\label{hyperplanes}
\end{eqnarray} 
\noindent
Consequently an admissible equilibrium configuration can exist only if  the $\epsilon_s$ have not all the same sign as already remarked at the end of section 2.
 Moreover, a unique equilibrium position  exists only  for $N=2$; setting $\mu=B_1-B_2$, we get:
 \begin{equation} 
 N^{0}_1=\frac{1}{\mu \epsilon_1};~~~~ N^{0}_2=-\frac{1}{\mu \epsilon_2}\label{equi2}
 \end{equation}
\noindent
For instance, in the case $N=3$,  we have a one parameter family of equilibrium solutions, reading ($0<\rho<1$)
\begin{equation}
\epsilon_1 N^{0}_1 =\frac{\rho}{B_1-B_3};\quad\; \epsilon_2N^{0}_2 = \frac{1-\rho}{B_2-B_3} ;
\quad \;\epsilon_3N^{0}_3=-(\epsilon_1 N^{0}_1 +\epsilon_2 N^{0}_2).\label{KerA}
\end{equation}
\noindent 
In (\ref{KerA}), in order the equilibrium  species populations be positive, we have to require 
$ (B_1-B_3)\epsilon_1$, and $(B_2-B_3)\epsilon_2$ to be positive quantities, while the ratios $\epsilon_1/\epsilon_3$ and $\epsilon_2/\epsilon_3$ have to be negative. 
\noindent
 We end the present subsection by remarking that in all cases, whether they are integrable or not, the N-species Volterra system enjoy a  sort of  box structure, being equipped with a number of  invariant submanifolds, obtaining when only a %(possibly not ordered)%
 subset of species is alive, this fact being dictated just by the initial conditions. So, for the two species case, the axes $N_1=0$ and $N_2=0$ are invariant submanifolds, for the three-species case we have the 6 invariant submanifolds given by the axes and by the planes $N_i =0$, and so forth. In particular the three-species case gives rise to three Lotka-Volterra systems.
%  \section{More on  the integrable case. The Volterra approach}\noindent
Once realized that integrability implies the non-uniqueness of the equilibrium configuration ($\forall$ $N>2$), we would like to stress that, for the special form of the matrix $A$ given by (\ref{matrix}), the integrals of motion are still functionally independent. This is readily seen as (\ref{H_r})  shows that the linear dependence of those integrals upon the momenta $p_r$ is in no-way affected by the specific form of the matrix $A$ (while the requirement that the $B_r$ be all distinct is mandatory!), so that the rank of the Jacobian matrix constructed with the gradients of the integrals of motion with respect to the canonical coordinates is maximal (namely $N$) whatever be that form. So, the integrable version of the $N$-species Volterra system is again a genuine Hamiltonian system with $N$ degrees of freedom. 
 \noindent
 Here we write down explicitly the expression of the Hamiltonian and of the integrals of motion in the integrable case, resuming what we sketched in formulas (\ref{H_r})-(\ref{A}). Denoting now by ${\mathcal H}_{int}$ the Hamiltonian (\ref{Ham})  we have:
 \begin{equation}
 {\mathcal H}_{int} = \sum_{r=1}^N \epsilon_r q_r-\exp[p_r-1+(\epsilon_r/2)\sum_{s=1}^N\epsilon_s(B_r-B_s)q_s]\label{H_int}
 \end{equation}
 \noindent
 and
 \begin{equation}
{\mathcal H}_r = p_r/\epsilon_r - (1/2)\sum_{s=1}^N\epsilon_s(B_r-B_s)q_s -t \quad r=1,\cdots,N\label{Hr_int}
\end{equation}
\noindent
so that 
\begin{equation}
{\mathcal H}_{rl}\equiv {\mathcal H}_r -{\mathcal H}_l =p_r/\epsilon_r -p_l/\epsilon_l -\frac{1}{2}(B_r-B_l)\sum_{s=1}^N \epsilon_s q_s, \quad s=1,\cdots, N.\label{Hrl}
\end{equation}
\noindent
The constants of motion (\ref{Hrl}) are mutually in involution. So we can take for instance $l=1$ and get $N-1$ independent integrals of motion in involution. The set can be completed by adding any function of the Hamiltonian, for instance the Hamiltonian  itself.
\noindent
The above formulas clearly show that, in the integrable case, both the Volterra Hamiltonian and the involutive family of integrals of motion depend on the quantities of life only through the
inner  products $(\epsilon,Q)$ and $(\epsilon,BQ)$, where $Q$ is  the vector of components $q_j$, while  by $BQ$ we have denoted the vector of components $B_j q_j$.
However, even in the completely integrable case, we did not succeed in reducing our problem to quadratures for $N$ larger than $2$, although this possibility is a  well known result in Classical Mechanics \cite{Arnold, Marsden}.
%Of course, if one comes back to the Lagrangian formulation, one will get  expressions involving the variables $q_s$ and $\dot q_s$. Let us also notice that
 It might be convenient to take as integrals of motion the exponentials of the quantities (\ref{Hr_int})
 \begin{equation}
 \exp(\epsilon_r {\mathcal H}_r )= \exp[p_r - \epsilon_r/2\sum_{s=1}^N \epsilon_s(B_r-B_s)q_s -\epsilon_r t]\label{exp}
 \end{equation}
 \noindent
hence choosing $\exp({\mathcal H_{rl}})$ as an alternative legitimate form for an involutive family of integrals of motion. In the simplest nontrivial case, $N=2$, (\ref{H_int}) reads (the subscript $V$ refers to Volterra):
  \begin{equation} 
 {\mathcal H}_V= \epsilon_1q_1 + \epsilon_2 q_2 - \exp[p_1+(1/2)\epsilon_1\epsilon_2(B_1-B_2)q_2] - \exp[p_2 - (1/2)\epsilon_1\epsilon_2(B_1-B_2)q_1]\label{Hv}
 \end{equation}
 \noindent
 The above formula can be slightly simplified through the canonical transformation (in fact, just a rescaling):
 \begin{equation}
 p_j  \to \tilde p_j = p_j/\epsilon_j; \quad q_j\to \tilde q_j = \epsilon_j q_j
 \end{equation}
 that maps (\ref{Hrl}) into:
 \begin{equation}
 {\mathcal H}_{rl} =\tilde p_r -\tilde p_l +(1/2)(B_r-B_l)\sum_j \tilde q_j
 \end{equation}
 \noindent
 and (\ref{Hv}) into:
 \begin{equation}
 { \mathcal H}_V = \tilde q_1 + \tilde q_2 - \exp[\epsilon_1(\tilde p_1+1/2)(B_1-B_2)\tilde q_2)] - \exp[\epsilon_2(\tilde p_2 - (1/2)(B_1-B_2)\tilde q_1]\label{newHv}
 \end{equation}
\noindent
%It might be worth to notice that in the integrable case the $q$-dependence of the Hamiltonian (\ref{H_int}) materializes just through the $global$ variables $Q :=\sum_s\tilde q_s$ and $BQ = \sum_s B_s\tilde q_s $. 
\medskip
\subsection{Integration  of  the  $N=2$ case  via  the  canonical  formalism and other Hamiltonian formulations.}
Let us slightly simplify the notations in (\ref{newHv}), by setting $\mu := B_1-B_2$, and introducing the new canonical variables:
\begin{equation}
P_1 = \frac{1}{\sqrt \mu}(\tilde p_1 +\frac{1}{2} \mu\tilde q_2); ~~~Q_1 =\frac{1} {\sqrt \mu} (-\tilde p_2 + \mu\frac{1}{2} \tilde q_1)
\end{equation}
\begin{equation}
P_2 = \frac{1}{\sqrt \mu}(\tilde p_1 -\mu\frac{1}{2} \tilde q_2); ~~~Q_2 = \frac{1}{\sqrt \mu} (\tilde p_2 +\ \frac{1}{2}\mu \tilde q_1)
\end{equation}
\noindent
In terms of these new  variables, the first integral:
\begin{equation}
{\mathcal H}_{12} =\tilde p_1 -\tilde p_2 - (\mu/2)(\tilde q_1 +\tilde q_2)
\end{equation}
takes the form
\begin{equation}
{\mathcal H}_{12} = \frac{1}{\sqrt \mu} (P_2-Q_2)\label{h12}
\end{equation}
\noindent
while the two-particle Hamiltonian reads:
\begin{equation}
{\mathcal H}_V = \frac{1}{\sqrt \mu}[Q_1 + Q_2 +P_1-P_2] -\exp{\sqrt \mu} (\epsilon_1 P_1) -\exp[-{\sqrt \mu}(\epsilon_2 Q_1)]
\end{equation}
\noindent
Inserting the  first integral (\ref{h12}), on the level surface ${\mathcal H}_{12}= C$, up to an irrelevant additive constant we can finally write:
\begin{equation}
{\mathcal H}_V= \frac{1}{\sqrt \mu}(Q_1+P_1) -\exp [{\sqrt \mu}(\epsilon_1 P_1)] -\exp[-{\sqrt \mu}(\epsilon_2 Q_1)]
 \end{equation}\noindent
It follows that, in terms of these new coordinates, the  above Hamiltonian  is a one-body Hamiltonian (integrable by definition), which is nothing but the traditional Lotka-Volterra Hamiltonian.
One may pervene to more elegant formulas by defining:
$$x \equiv \exp {\sqrt \mu}(\epsilon_1 P_1) ;~~~ y\equiv \exp[-{\sqrt \mu}(\epsilon_2 Q_1)]$$
leading to:
\begin{eqnarray}
\dot x = \epsilon_1x(1 - \mu \epsilon_2y) \label{predatorpreya}\\
\dot y =  \epsilon_2 y (1+\mu\epsilon_1 x)\label{predatorpreyb}
\end {eqnarray}
The (nontrivial and stable) equilibrium position is the pair $(x^{(0)}=-\frac{1}{\mu\epsilon_1}, y^{(0)}=\frac{1}{\mu\epsilon_2})$
whence it follows that it can belong to the first quadrant only if the coefficients of spontaneous growth have opposite sign, as it is natural if we require that the predator species can survive eating the prey one. 
\noindent
As well known \cite{Volterra2, Braun},  the equations for the orbits of (\ref{predatorpreya}) and (\ref{predatorpreyb}) can be written in closed form:
\begin{equation}
(x \exp(\mu\epsilon_1x))^{\epsilon_2}~=~K(y\exp(-\mu\epsilon_2y))^{\epsilon_1}\label{orbits}
\end{equation}
where $K$ is a positive constant.
For the sake of completeness, we recall here the standard form of the predator-prey equations (keeping skew-symmetry) and of  their hamiltonian formulation, that might be useful to compare with that described at the beginning of this subsection.
Denoting by $x$ and $y$ the two species, we have.
\begin{eqnarray}
\dot x= \epsilon_1 x - axy\label{predatorpreya}\\
\dot y =\epsilon_2 y +axy\label{predatorpreyb}  
\end{eqnarray}
\noindent
which coincide with (\ref{predatorpreya}) and (\ref{predatorpreyb}) if we set $a=\mu\epsilon_1\epsilon_2$
There are two equiilbrium positions: the trivial one $(0,0)$, and the center $(\epsilon_1/a,-\epsilon_2/a)$. Assuming $a>0$ (and thus $\mu<0$), it belongs to the first quadrant provided 
$\epsilon_1>0,~\epsilon_2<0$.
The invariant curve, that defines the family of orbits and plays also the role of Hamiltonian, reads:
\begin{equation}
h(x,y)=\epsilon_2\ln x -\epsilon_1 \ln y +a(x+y)\label{invcur}
\end{equation}
\noindent
The equations (\ref{predatorpreya}),(\ref{predatorpreyb}) can be cast in the following Hamiltonian form:
\begin{eqnarray}
\dot x = -xy\frac{\partial h}{\partial y}\label{hprpra}\\
\dot y = xy \frac{\partial h}{\partial x}\label{hprprb}
\end{eqnarray}
\noindent
which involves the Poisson matrix
\begin{equation}
P=\begin{pmatrix}
0&-xy\\
xy&0\label{poisson}
\end{pmatrix}
\end{equation}
\noindent
The simple change of variables:
$x=\exp(\tilde x),~y=\exp(\tilde y)$ 
\noindent
transforms (\ref{predatorpreya},\ref{predatorpreyb}) into:
\begin{eqnarray}
\dot {\tilde x} = \epsilon_1 -a\exp(\tilde y)\label{newLVa}\\
\dot {\tilde y} = \epsilon_2 + a\exp(\tilde x)\label{newLVb}
\end{eqnarray}
the Hamiltonian (\ref{invcur}) into:
\begin{equation}
\tilde h(\tilde x,\tilde y)= \epsilon_2 \tilde x -\epsilon_1 \tilde y +a(\exp(\tilde x) + \exp(\tilde y))\label{newinv}
\end{equation}
\noindent
and the Poisson matrix (\ref{poisson}) into the canonical one
\begin{equation}
J=\begin{pmatrix}
0&-1\\
1&0\label{canonical}
\end{pmatrix}
\end{equation}
\subsection{The $N$-species system: integration through the Volterra's approach.}
Here we briefly recall the procedure followed by Volterra to  integrate the N-species system in terms of the  natural coordinates, i.e.the population numerosities. In the following, we do not preclude the possibility to have an odd number of species, so $N$ can be even or odd.
\noindent
Volterra defines the quantities ${\mathcal N} := \sum_s \epsilon_sN_s$ and ${\mathcal M}:=1-\sum_s \epsilon_s B_s N_s$, then rewriting the original dynamical system (1) as:
 \begin{equation*}
 \dot N_r = \epsilon_r N_r( 1+\sum_s(B_r-B_s)\epsilon_sN_s)\label{Volt2}
 \end{equation*}
 \noindent
 or, in other terms:
  \begin{equation*}
 \dot N_r = \epsilon_r N_r( B_r {\mathcal N} + {\mathcal M})\label{Volt3}
 \end{equation*}
  \noindent
  namely:
   \begin{equation*}
   (1/\epsilon_r) d/dt \log N_r = ( B_r {\mathcal N} + {\mathcal M})
   \end{equation*}
Note that the previous equation implies $\sum_r(\dot N_r - \epsilon_rN_r) =0$, which is just the conservation of the Hamiltonian (\ref{Ham}). It is evident that the involutivity constraints on the coefficients $A_{rs}$ entail a typical  Mean Field dynamics. Each species interacts with the others through the collective~variables ${\mathcal N}$ and ${\mathcal M}$. By taking two different values of the index $r$ and subtracting, the variable $ {\mathcal M}$ can be eliminated, resorting to:
  \begin{equation*}
  \frac{(1/\epsilon_r) d/dt \log N_r -(1/\epsilon_s) d/dt \log N_s}{B_r-B_s} = {\mathcal N}, 
  \end{equation*}
  \noindent
 that yields $N-2$ integrals of motion. Indeed, in term of the variables
  \begin{equation}
  Y_k \equiv {(1/\epsilon_k})\log N_{k}\label {Y},
  \end{equation}
\noindent
we get the linear formula
 \begin{displaymath}
\frac{1}{B_r-B_j}(Y_r-Y_j)-\frac{1}{B_s-B_j}(Y_s-Y_j) = C_{rs}.
\end{displaymath}
\noindent
  \noindent 
  The simplest non-trivial case is the three-species case, where an elementary calculation shows that the three equations above are indeed the same, yielding:
  \begin{equation}
 ( B_2-B_3)Y_1 + (B_3-B_1)Y_2 + (B_1-B_2)Y_3 = const.\label{Third integral}
  \end{equation} 
  \noindent
 We remind the form of  the  equations of motion for the integrable three species case: 
 \begin{eqnarray}
 \dot N_1 = \epsilon_1 N_1 + \epsilon_1\epsilon_2\ (B_1-B_2)N_1 N_2 +
  \epsilon_1\epsilon_3(B_1-B_3)N_1 N_3\label{first}\\
 \dot N_2 = \epsilon_2 N_2+ \epsilon_2\epsilon_1(B_2-B_1)N_2 N_1 + 
 \epsilon_2 \epsilon_3 (B_2 -B_3)N_2 N_3\label{second}\\
 \dot N_3 =  \epsilon_3 N_3 +\epsilon_3\epsilon_1(B_3-B_1)N_3 N_1 + \epsilon_3\epsilon_2(B_3-B_2)N_3N_2\label{third}
 \end{eqnarray}
and add the expression of the integrals of motion for the system (\ref {first})-(\ref{third})
 in terms of the numerosities and of the quantities of life.
Taking into account (\ref{Y}), formula  ($ \Ref{Third integral}$) takes the form
  \begin{equation}
  N_1^{(B_2-B_3)/\epsilon_1}N_2^{(B_3-B_1)/\epsilon_2}N_3^{(B_1-B_2)/\epsilon_3} = I_{123}\label{thirdnew}
  \end{equation}
  Of course, (\ref{thirdnew}) has a meaning only for nonzero initial data, where it can be written as well as:
 \begin{equation}
  (N_1/N_1{(0)})^{(B_2-B_3)/\epsilon_1}(N_2/N_2{(0)})^{(B_3-B_1)/\epsilon_2}(N_3/N_3{(0)})^{(B_1-B_2)/\epsilon_3} = 1 \label{thirdneww}
  \end{equation}
\noindent
  The Hamiltonian ${\mathcal H}$ is written as: 
  \begin{equation}
  {\mathcal H} =\sum_{r=1}^N \epsilon_r\int_0^tN_r(t^\prime)dt^\prime -N_r \label{Hamecol}
  \end{equation}
  \noindent
   implying that its value equals minus the total population at $t=0$.
   \noindent
 The involutive integrals $\exp({\mathcal H_r})$ read:
  \begin{equation}
 \exp( {\mathcal H_r})=N_r\exp[ -\sum_s(B_r-B_s)\epsilon_r\epsilon_s\int_0^tN_s(t^\prime)dt^\prime ]
  \end{equation}
 \noindent
  so each of them equals the corresponding initial population.
In the next section, in which we will give both analytical and numerical results, we will focus attention on the three species integrable case, 
which escapes the general analysis presented by Volterra in \cite{Volterra2}, focussed on the study of an even number of species, mostly assuming invertibility of the interaction matrix $A$. To our knowledge, Volterra copes with the three species problem  only in the case of null spontaneous growth coefficients, where  he proves the existence of periodic orbits. By the way, a full treatment of the integrable three species case can be found in \cite{Baigent}, denoted as ``A three species food chain".
 \section{Novel numerical and analytical results: from 3 to $N$.}
 \noindent
In this section we propose a way to construct the integrals of motion for the model under scrutiny which is alternative  to the one followed by Volterra,  recalled in the previous section and in \cite {RS}. In the first subsection we present the case $N=3$, by giving also a numerical  integration of the equations of motion (\ref{first})-(\ref{third}) for different choices of the relevant parameters. In the second subsection we extend the construction to the generic $N$-species case.
\subsection{The three species case.}
We consider the case $N=3$ here. To simplify a little bit the notations, let us set:
 \begin{equation}
B_1-B_2 = \alpha; ~~ B_2-B_3 = \beta
\end{equation}
\noindent 
whence $B_1-B_3 =\alpha +\beta$.
Recalling (\ref{KerA}), we notice that, if $\epsilon_1, \epsilon_2$ are positive quantities, we have to require:
\begin{equation}
 \beta>0,~ \alpha +\beta>0\label{alfabeta}
 \end{equation}
 \noindent
 implying that, if $\alpha$ is negative, its absolute value has to be less than $\beta$.
\noindent
 We start from the equations of motion
\begin{eqnarray}
 \dot N_1 = \epsilon_1 N_1 + \epsilon_1\epsilon_2 \alpha N_1 N_2 +
 \epsilon_1\epsilon_3(\alpha+\beta)N_1 N_3\label{ffirst}\\
\dot N_2 = \epsilon_2 N_2- \epsilon_2\epsilon_1 \alpha N_2 N_1 + 
 \epsilon_2 \epsilon_3 \beta N_2 N_3\label{ssecond}\\
 \dot N_3 =  \epsilon_3 N_3 -\epsilon_3\epsilon_1(\alpha+\beta)N_3 N_1 -\epsilon_3\epsilon_2\beta N_3N_2\label{tthird}\
 \end{eqnarray}
  \noindent
 and look for an integral of motion written as
\begin{equation}\label{const}
e^{A(N_1(t)+N_2(t)+N_3(t))}N_1(t)^mN_2(t)^nN_3(t)^k,
\end{equation}
for suitable constants $A,m,n,k$. By deriving (\ref{const}) with respect to time, dividing by (\ref{const}) itself and making use of (\ref{first})-(\ref{third}) we find 
\begin{equation}\begin{split}
&\epsilon_1N_1\left(A-n\alpha \epsilon_2-k(\alpha+\beta)\epsilon_3\right)+\epsilon_2N_2\left(A+m\alpha \epsilon_1-k\beta\epsilon_3\right)+\\
&+\epsilon_3N_3\left(A+n\beta \epsilon_2+m(\alpha+\beta)\epsilon_1\right)+\epsilon_1 m+\epsilon_2 n+\epsilon_3 k=0.
\end{split}\end{equation}
The coefficients of the $N_i's$, $i=1,2,3$, are all compatible each other if $\epsilon_1 m+\epsilon_2 n+\epsilon_3 k=0$, giving e.g. $A=k\beta\epsilon_3-m\alpha \epsilon_1$. So, the following quantity
\begin{equation}\label{const1}
e^{\left(k\beta\epsilon_3-m\alpha \epsilon_1\right)\left(N_1(t)+N_2(t)+N_3(t)\right)}N_1(t)^mN_2(t)^nN_3(t)^k=I_{m,n,k}
\end{equation}
is a conserved quantity if the parameters $(m,n,k)$ satisfies $\epsilon_1 m+\epsilon_2 n+\epsilon_3 k=0$. Notice that (\ref{thirdnew}) can be rewritten as:
\begin{equation}
 \frac{ N_1^{\beta/\epsilon_1}N_3^{\alpha/\epsilon_3}}{N_2^{(\alpha+\beta)/\epsilon_2}} = I_{123},\label{thirdnew1}
  \end{equation}
  \noindent
  This conserved quantity, that  corresponds to the choice $(m,n,k)=(\beta/\epsilon_1,-(\alpha+\beta)/\epsilon_2,\alpha/\epsilon_3)$ coincides with that of formula (\ref{thirdneww}).
Of the three parameters $(m,n,k)$, only two are independent because of the relation $\epsilon_1 m+\epsilon_2 n+\epsilon_3 k=0$. It is possible to write two different surfaces by choosing properly the values of the constants $m,n,k$: if these surfaces intersect by defining a closed curve, the corresponding motion defined by (\ref{first}-\ref{third}) will be periodic. Let us make an example. If we take
$$
\alpha=1, \;\beta=2\;\epsilon_1=\epsilon_2=1,\; \epsilon_3=-1.
$$
Then, two conserved quantities are given by
\begin{equation}\label{Ic12}
\frac{N_1^2}{N_2^3N_3}=I_1,\quad \quad e^{-2(N_1+N_2+N_3)}N_2N_3=I_2.
\end{equation}
We give the plot of the closed orbit $(N_1(t),N_2(t),N_3(t))$ numerically obtained by taking the initial conditions $N_1(0)=N_2(0)=N_3(0)=1$, the plot of $I_1$, of $I_2$ and of the closed orbit all together: as it can be seen from the last figure the two surfaces $I_1$ and $I_2$ intersect in a closed curve and the motion is constrained on this curve.
\begin{figure}[H]
\centering
\includegraphics[scale=0.7]{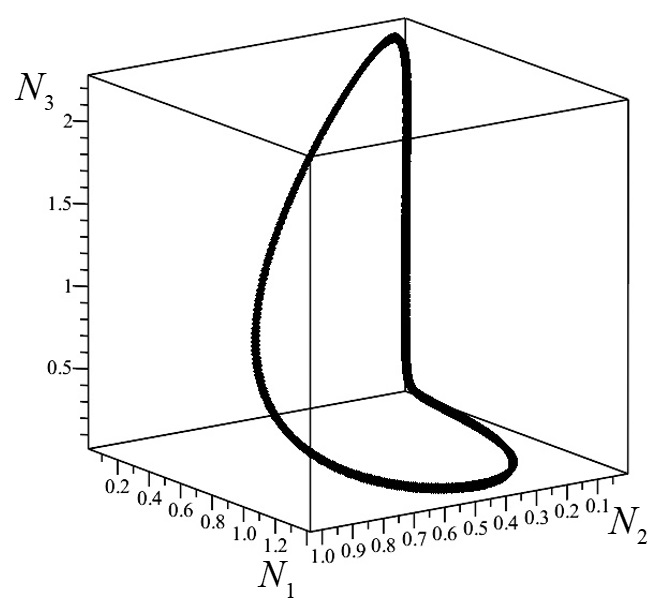}
\caption{Plot of the closed  orbit $(N_1(t),N_2(t),N_3(t))$ corresponding to the initial conditions $N_1(0)=N_2(0)=N_3(0)=1$}
\end{figure}
%\begin{figure}[H]
%\centering
%\includegraphics[scale=0.7]{1.jpg}
%\caption{Plot of the surface $I_1$ in (\ref{Ic12}) corresponding to $N_3=\frac{N_1^2}{N_2^3}$.}
%\end{figure}
%\begin{figure}[H]
%\centering
%\includegraphics[scale=0.8]{10.jpg}
%\caption{Plot of the surfaces  $I_2$ in (\ref{Ic12}) corresponding to $e^{-2(N_1+N_2+N_3)}N_2N_3=e^{-6}$.}
%\end{figure}
\begin{figure}[H]
\centering
\includegraphics[scale=0.7]{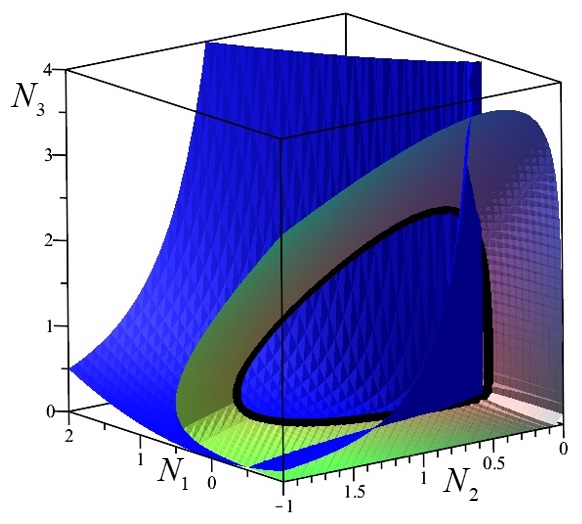}
\caption{Plot of  $N_3=\frac{N_1^2}{N_2^3}$ and $e^{-2(N_1+N_2+N_3)}N_2N_3=e^{-6}$ with the numerically obtained closed orbit (in black)}
\end{figure}
  \noindent
  Together with the geometrical picture given by the orbits, we have the dynamical picture associated
  with the time-behavior of the system.
  \noindent
We will consider two different cases of periodic behavior, one with $\alpha >0$, and the other with $\alpha<0$, with different initial conditions.
 \noindent
Case 1: $\epsilon_1=\epsilon_2=1; \epsilon_3=-1; \alpha=0.5; \beta =2.5$.
$$N_1{(0)} = 1,  N_2{(0)} = 1, N_3{(0)}= 1, t\in[0, 30]$$
\begin{eqnarray}
\dot N_1 = N_1 + 0.5 N_1 N_2 - 3 N_1N_3, \\
\dot N_2  = N_2 - 0.5 N_1N_2 - 2.5 N_2 N_3 \\
\dot N_3 = - N_3 + 3 N_1 N_3 + 2.5 N_3 N_2\\
 \end{eqnarray}
\noindent The corresponding plots of $N_i$, $i=1,...,3$ are given in figure (\ref{figna}).
\noindent
Case 2: $\epsilon_1=\epsilon_2=1; \epsilon_3=-1,\alpha=-1/2,\beta =1$.
$$ N_1{(0)} = 1,  N_2{(0)} = 1/2, N_3{(0)}= 1/2, t\in[0, 30]$$
$$\dot N_1 = N_1 - 0.5 N_1 N_2 -  N_1N_3$$
$$\dot N_2 = N_2 +0.5 N_1N_2 - 0.5 N_2 N_3 $$
$$\dot N_3  = - N_3 +N_1 N_3 + 0.5 N_3 N_2$$
\noindent The corresponding plots of $N_i$, $i=1,...,3$ are given in figure (\ref{fignb}).
\begin{figure}[H]
\centering
\includegraphics[scale=0.65]{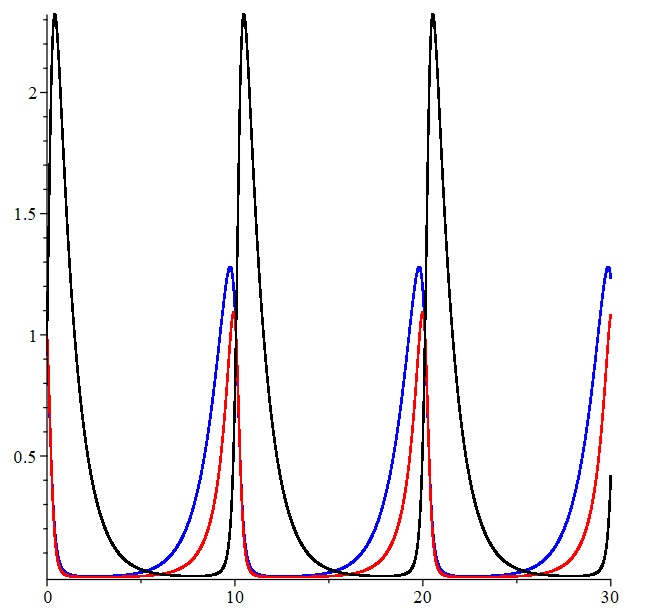}
\caption{Oscillations of $N_1$ (in red), $N_2$ (in blue) and $N_3$ (in black) corresponding to the initial conditions $N_1(0) =1, N_2(0)=1, N_3(0)=1$ $\alpha=1/2$, $\beta=5/2$.}
\label{figna}
\end{figure}
\begin{figure}[H]
\centering
\includegraphics[scale=0.65]{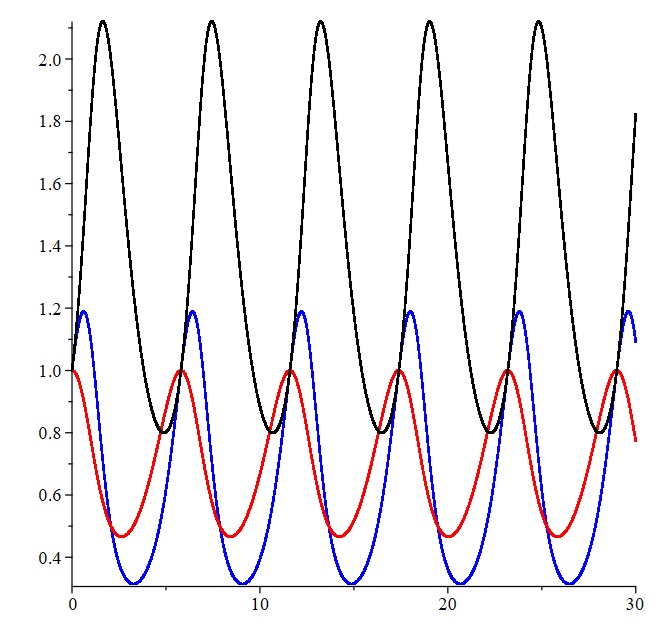}
\caption{Oscillations of $N_1$ (in red), $N_2$ (in blue) and $N_3$ (in black) corresponding to the initial conditions $N_1(0) =1, N_2(0)=1, N_3(0)=1$ $\alpha=-1/2$, $\beta=1$.}
\label{fignb}
\end{figure}
%\begin{figure}
%\centering
%\includegraphics[scale=0.6]{N1caso3.jpg}[H]
%\caption{Oscillations of the $N_1$ with initial conditions $N_1(0)=1, N_2(0)=1,N_3(0)=1$,$\alpha=-0.5$, $\beta=-0.5$}
%\end{figure}
%\noindent
%We now exhibit three pictures associated wih the same values of $\alpha,\beta$, namely $\alpha=0.5,\beta=2.5$, but with different values of the initial conditions.
%\begin{figure}
%\centering
%\includegraphics[scale=0.6]{N2cond111.jpg}[H]
%\caption{Oscillations of $N_2$ with initial conditions $N_1(0)=1$,$N_2(0)=1/2$, $N_3(0)=1/2$,$\alpha=0.5$, $\beta=2.5$}
%\end{figure}
%\begin{figure}[H]
%\centering
%\includegraphics[scale=0.6]{N1nons.jpg}
%\caption{Oscillations of $N_1$,$N_3$ species with initial conditions $N_1(0)=1$,$N_2(0)=0$, $N_3(0)=1$, $\alpha=0.5$, $\beta=2.5$}
%\end{figure}
%\noindent
To conclude this subsection it sounds appropriate comparing the fully nonlinear picture described above with the one corresponding to the linearized evolution in the neighbourhood of (one suitable point) of the equilibrium configuration. 
To this aim, let us set
\begin{equation} 
N_i(t)=N_i^{0}+x_i(t), \quad i=1,2,3
 \end{equation}
\noindent
where $x_i$ are small quantities. At first order in the $x_i$, the set of equations (\ref{model}) becomes
\begin{equation}\begin{split}
&\dot x_1 = \frac{B_1-B_2}{B_1-B_3}\rho\epsilon_{2}x_2+\rho\epsilon_3 x_3,\\
&\dot x_2 = \frac{B_2-B_1}{B_2-B_3}(1-\rho)\epsilon_{1}x_1+(1-\rho)\epsilon_3 x_3,\\
&\dot x_3 =\left(\frac{B_1-B_3}{B_2-B_3}+\rho\frac{B_2-B_1}{B_2-B_3}\right)\epsilon_1x_1+\left(1-\rho\frac{B_1-B_2}{B_1-B_3}\right)\epsilon_2x_2.\label{linearsystem}
\end{split}\end{equation}
We need the eigenvalues of the matrix of the coefficients, given by
\begin{equation}
M=\begin{pmatrix}
0 & \frac{B_1-B_2}{B_1-B_3}\rho\epsilon_{2} & \rho\epsilon_3\\
\frac{B_2-B_1}{B_2-B_3}(1-\rho)\epsilon_{1} & 0 & (1-\rho)\epsilon_3\\
\left(\frac{B_1-B_3}{B_2-B_3}+\rho\frac{B_2-B_1}{B_2-B_3}\right)\epsilon_1&\left(1-\rho\frac{B_1-B_2}{B_1-B_3}\right)\epsilon_2&0
\end{pmatrix}
\end{equation}
The characteristic polynomial is given by:
\begin{equation}
\lambda^3-\lambda^2\textrm{Tr}(M)+\frac{\lambda}{2}\left(\textrm{Tr}(M)^2-{Tr}(M^2)\right)-\textrm{De}t(M)=0,
\end{equation}
and, since $\textrm{Tr}(M)$ and $\textrm{Det}(M)$ are both equal to zero, it reduces to
\begin{equation}
\lambda(\lambda^2-\frac{1}{2}\textrm{Tr}(M^2))=0.
\end{equation}
We see that an eigenvalue is always zero, the sum of the other two is zero. So the three eigenvalues are $(0,\lambda_1,-\lambda_1)$. If $\lambda_1$ is real, then the fixed point is unstable. If $\lambda_1$ is imaginary (it happens if $\textrm{Tr}(M^2)<0$) then the matrix $M$ must have three independent eigenvectors in order the point to be stable \cite{Braun}. The eigenvector corresponding to the eigenvalue $0$ is $(\frac{B_2-B_3}{\epsilon_1}, \frac{B_3-B_1}{\epsilon_2}, \frac{B_1-B_2}{\epsilon_3})$ whereas the other two must be independent. The point is not asymptotically stable obviously. So, if $\textrm{Tr}(M^2)<0$ one has periodic orbits, the period being $\textrm{T}=\frac{2\sqrt{2}\pi}{\sqrt{|\textrm{Tr}(M^2)|}}$.
\noindent
By setting $B_1=B_2+\alpha$ and $B_2=B_3+\beta$, we get:
$$\textrm{Tr}(M^2)=2\epsilon_3\left(\epsilon_2+\rho(\epsilon_1-\epsilon_2)\right)-2\alpha\rho(1-\rho)\frac{\epsilon_1(\epsilon_2-\epsilon_3)\alpha+\epsilon_3(\epsilon_2-\epsilon_1)\beta}{\beta(\alpha+\beta)}$$
Recalling that, in all cases we have considered, we always choose $\epsilon_1=1$ $\epsilon_2=1$, $\epsilon_3=-1$, the above expression becomes:
$$-2-4\frac{\alpha^2\rho(1-\rho)}{\beta(\beta+\alpha)}$$
which is surely negative, $\forall  0<\rho <1$, provided $\frac{\alpha}{\beta} \in (-1,+\infty)$.
\noindent
Once fulfilled, the condition will hold for any point on the equilibrium line.
\noindent
Finally, let us compare the behaviour of the nonlinear model and its linearization, given in subsection (\ref{sub3.3}) (see equations (\ref{linearsystem}) and the discussion after these equations). 
We make two numerical examples by taking the following values of the parameters:
\begin{equation}
\epsilon_1=\epsilon_2=1,\;\; \epsilon_3=-1,\;\; \alpha=1,\;\; \beta=2.\;\;
\end{equation} 
The equilibrium positions are given by
\begin{equation}
N_1^0=\frac{\rho}{3},\;\; N_2^0=\frac{1-\rho}{2}, \;\;N_3^0=\frac{1}{2}-\frac{\rho}{6}.
\end{equation}
We start with an initial condition close to the equilibrium point $(N_1^0,N_2^0,N_3^0)=(\frac{1}{6},\frac{1}{4},\frac{5}{12})$ corresponding to $\rho=\frac{1}{2}$: we choose $(N_1(0),N_2(0),N_3(0))=(\frac{1}{6}+0.1,\frac{1}{4},\frac{5}{12})$. The trajectory determined by the numerical solution of the non-linear system (\ref{model}) and the one determined by the analytical solution of the set of equation (\ref{linearsystem}) are compared in figure (\ref{figu5}). Also, the evolution of the population $N_1(t)$ and the evolution of $N_1^0+x_1(t)$ is given in figure (\ref{figu6})
\begin{figure}[H]
\centering
\includegraphics[scale=0.65]{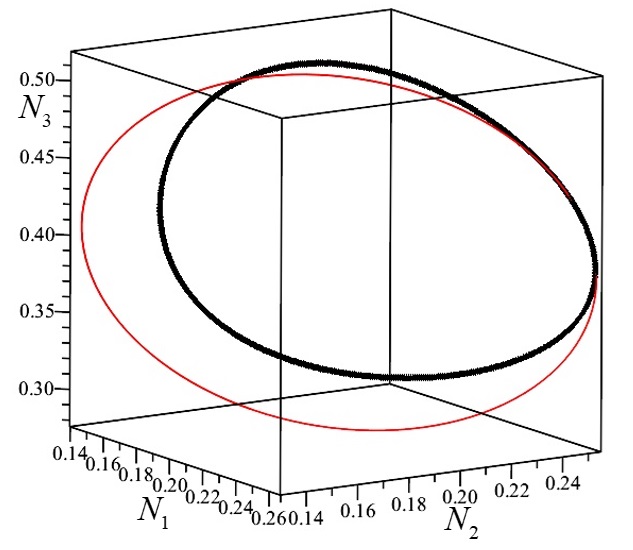}
\caption{Plot of the closed  orbit $(N_1(t),N_2(t),N_3(t))$ for the system (\ref{model}) corresponding to the initial conditions $N_1(0)=\frac{1}{6}+0.1$, $N_2(0)=\frac{1}{4}$,$N_3(0)=\frac{5}{12}$ with $\epsilon_1=\epsilon_2=1$, $\epsilon_3=-1$, $ \alpha=1$, $\beta=2$ (in black) and of $(\frac{1}{6}+x_1(t), \frac{1}{4}+x_2(t), \frac{5}{12}+x_3(t))$, where $(x_1(t),x_2(t),x_3(t))$ are solution of the system (\ref{linearsystem}) with initial conditions $x_1(0)=0.1$, $x_2(0)=0$, $x_3(0)=0$ (in red).}\label{figu5}
\end{figure}
\begin{figure}[H]
\centering
\includegraphics[scale=0.65]{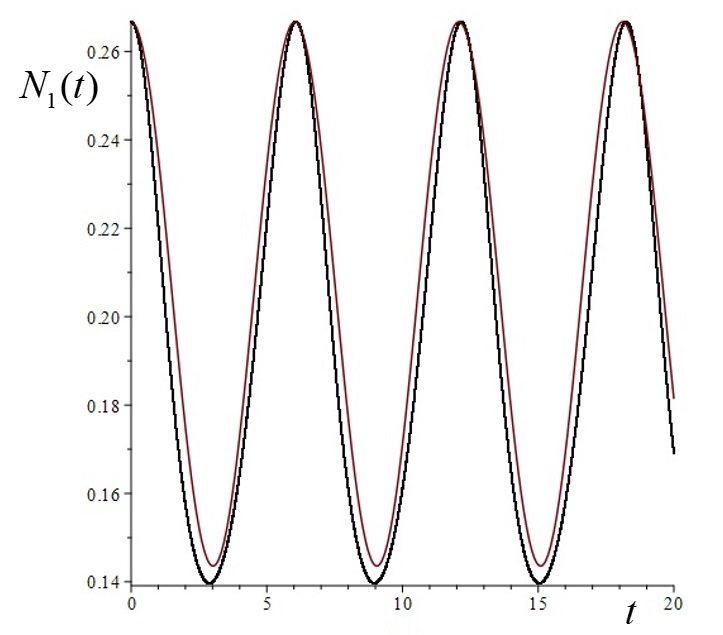}
\caption{Plot of the evolution of $N_1(t)$ for the system (\ref{model}) corresponding to the initial conditions $N_1(0)=\frac{1}{6}+0.1$, $N_2(0)=\frac{1}{4}$,$N_3(0)=\frac{5}{12}$ with $\epsilon_1=\epsilon_2=1$, $\epsilon_3=-1$, $ \alpha=1$, $\beta=2$ (in black) and of $\frac{1}{6}+x_1(t)$, where $x_1(t)$ is the solution of the system (\ref{linearsystem}) with initial conditions $x_1(0)=0.1$, $x_2(0)=0$, $x_3(0)=0$ (in red).}\label{figu6}
\end{figure}
Now we take an initial condition closer to the equilibrium point $(N_1^0,N_2^0,N_3^0)=(\frac{1}{6},\frac{1}{4},\frac{5}{12})$ with respect to the previous one. We choose $(N_1(0),N_2(0),N_3(0))=(\frac{1}{6}+0.01,\frac{1}{4},\frac{5}{12})$. Again, the trajectory determined by the numerical solution of the non-linear system (\ref{model}) and the one determined by the analytical solution of the set of equation (\ref{linearsystem}) are compared in figure (\ref{figu7}). Also, the evolution of the population $N_1(t)$ and the evolution of $N_1^0+x_1(t)$ is given in figure (\ref{figu8}). As it can be seen by the figures the trajectories are really close. Also, the period of the linear system $\textrm{T}=\frac{2\sqrt{2}\pi}{\sqrt{|\textrm{Tr}(M^2)|}}$ is a very good approximation for the period of the system in these cases.
\begin{figure}
\centering
\includegraphics[scale=0.7]{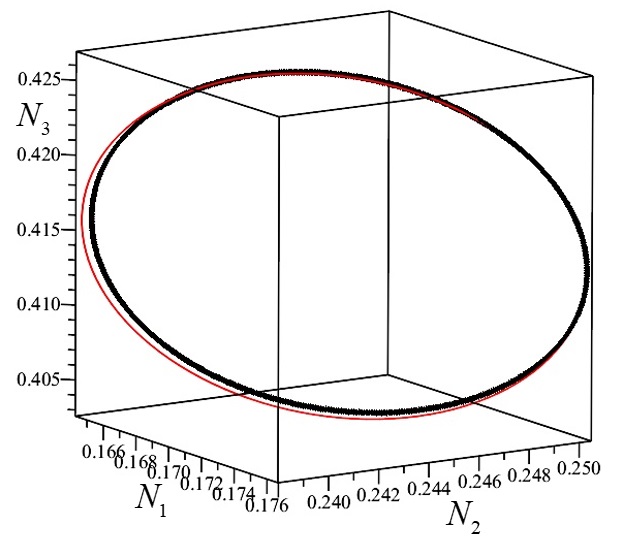}
\caption{Plot of the closed  orbit $(N_1(t),N_2(t),N_3(t))$ for the system (\ref{model}) corresponding to the initial conditions $N_1(0)=\frac{1}{6}+0.01$, $N_2(0)=\frac{1}{4}$,$N_3(0)=\frac{5}{12}$ with $\epsilon_1=\epsilon_2=1$, $\epsilon_3=-1$, $ \alpha=1$, $\beta=2$ (in black) and of $(\frac{1}{6}+x_1(t), \frac{1}{4}+x_2(t), \frac{5}{12}+x_3(t))$, where $(x_1(t),x_2(t), x_3(t))$ are solution of the system (\ref{linearsystem}) with initial conditions $x_1(0)=0.01$, $x_2(0)=0$, $x_3(0)=0$ (in red).}\label{figu7}
\end{figure}
\begin{figure}
\centering
\includegraphics[scale=0.7]{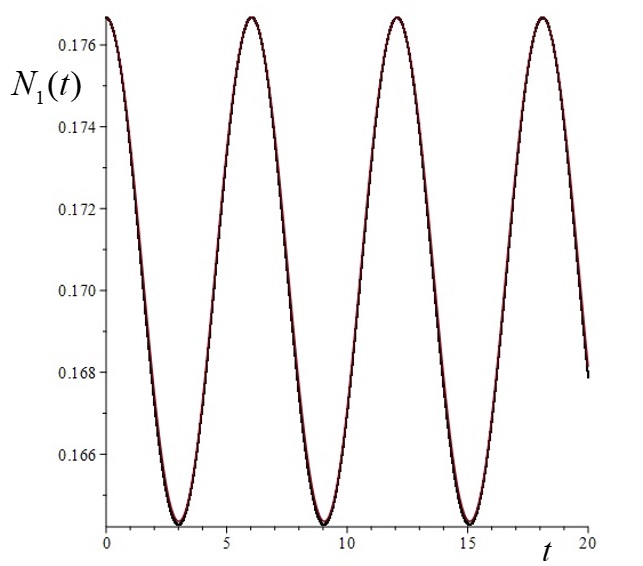}
\caption{Plot of the evolution of $N_1(t)$ for the system (\ref{model}) corresponding to the initial conditions $N_1(0)=\frac{1}{6}+0.01$, $N_2(0)=\frac{1}{4}$,$N_3(0)=\frac{5}{12}$ with $\epsilon_1=\epsilon_2=1$, $\epsilon_3=-1$, $ \alpha=1$, $\beta=2$ (in black) and of $\frac{1}{6}+x_1(t)$, where $x_1(t)$ is the solution of the system (\ref{linearsystem}) with initial conditions $x_1(0)=0.01$, $x_2(0)=0$, $x_3(0)=0$ (in red).}\label{figu8}
\end{figure}
The agreement between the exact (numeric) nonlinear picture and the linearized (analytic) one looks extremely, and somehow astonishingly, good. The global or local character of this agreement will be discussed in a separate work.
\subsection{Extension to $N$ species.}
\begin{figure}[H]
\centering
\includegraphics[scale=0.65]{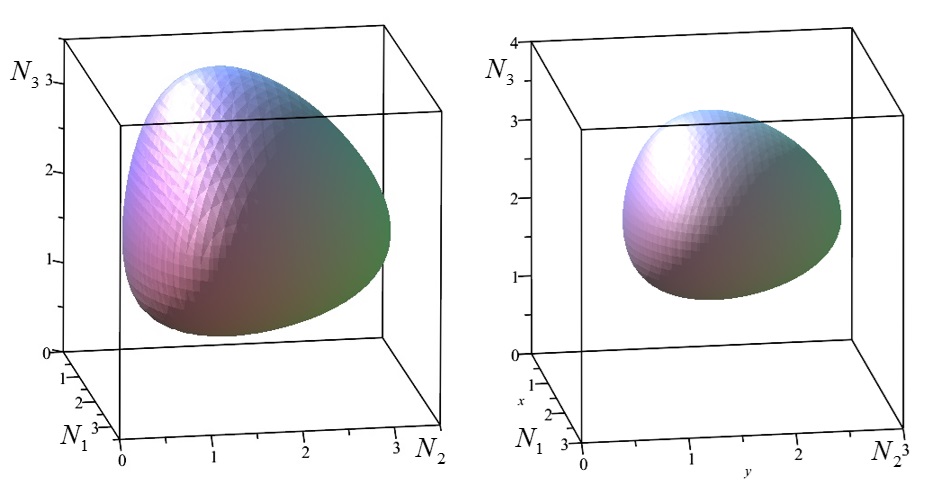}
\caption{Plot of $\frac{N_1^{2}}{e^{2 N_1}}\frac{N_2^{3}}{e^{2 N_2}}\frac{N_3^{4}}{e^{2 N_3}}=0.001$ (left) and of the projection of  $\frac{N_1^{2}}{e^{2 N_1}}\frac{N_2^{3}}{e^{2 N_2}}\frac{N_3^{4}}{e^{2 N_3}}=0.0005$ in the space $(N_1, N_2, N_3)$ (right). The value of $N_4$ has been fixed to be equal to 2.}
\label{fig4}
\end{figure}

We recall that the integrable case of the Volterra equations with $N$ species reads
\begin{equation}
\frac{d N_r}{dt} = \epsilon_r N_r + \sum_{s\ne r=1}^N A_{rs}N_rN_s \label{Volt1}
\end{equation}
where the matrix of the interactions s taken as
\begin{equation}
A_{rs} =\epsilon_r \epsilon_s (B_r-B_s)~~~r,s=1,\cdots N	\label{invo}	
\end{equation}			
Let us look for an integral of motion of the type:
\begin{equation}\label{constgen}
e^{A\sum_i N_i(t)}\prod_{i=1}^N N_i(t)^{w_i},
\end{equation}
where $A$ and $w_i$ are $N+1$ suitable constants. By requiring that the derivative of (\ref{constgen}) with respect to time is zero and by using equations (\ref{Volt1}), we get, after some manipulations:
\begin{equation}\label{aeq}
\sum_{k=1}^N \epsilon_k w_k =0, \quad\quad A=-\sum_{k=1}^N B_k\epsilon_k w_k,
\end{equation}
so that the quantities 
\begin{equation}\label{eqnew}
e^{-\sum_{k=1}^N B_k\epsilon_{k}w_k\left(\sum_i N_i(t)\right)}\prod_{i=1}^N N_i(t)^{w_i}=I_{1,..,N},
\end{equation}
are a parametric family of conserved quantities, depending on $N-1$ parameters since the constraint $\sum_{k=1}^N \epsilon_k w_k =0$ must be satisfied. Notice that equation (\ref{eqnew}) can be written as the following product of functions of a single variable:
\begin{equation}\label{eqn1}
\prod_{i=1}^N \frac{N_i^{w_i}}{e^{A N_i}} = I_{1,..,N}
\end{equation}
where $A$ is defined in (\ref{aeq}). The function $f(x)=x^w/e^{ax}$ plays a crucial role in a quite simple proof that the orbits of the original two-populations Volterra model is periodic (see e.g. \cite{Braun}). Indeed, for $x$ positive, it has a maximum in $x=w/a$ and decreases to zero by going to $x=0$ and $x=\infty$, similar to a gaussian bell shaped curve. Let us look at equation (\ref{eqn1}) supposing to vary just the value of the parameter $I_{1,..,N}$ (the other parameters being fixed):
\begin{equation}\label{prodeq}
f_1(N_1)\cdot f_2(N_2)\cdot \ldots \cdot f_{N}(N_N)=I_{1,...,N}, \quad f_{i}(N_i)\doteq \frac{N_i^{w_i}}{e^{A N_i}}
\end{equation}
Each of the functions $f_i(N_i)$ has a maximum in $N_i=w_i/A$ equal to $M_i=\left(\frac{w_i}{A e}\right)^{w_{i}}$: it follows that if 
\begin{equation}
I_{1,...,N}>\prod_{i=1}^N M_i
\end{equation}
then the equation (\ref{prodeq}) has no real solution. Also, if  
\begin{equation}
I_{1,...,N}=\prod_{i=1}^N M_i
\end{equation}
then equation (\ref{prodeq}) has just one real solution, given by $N_i=w_i/A$, $i=1,...,N$. Let us suppose now that  
\begin{equation}
I_{1,...,N}<\prod_{i=1}^N M_i.
\end{equation}
We set
\begin{equation}
I_{1,...,N}=\lambda \prod_{i=2}^N M_i, \quad \lambda <M_1
\end{equation}
and look at the equation 
\begin{equation}\label{prodeq1}
f_1(N_1)\cdot f_2(N_2)\cdot \ldots \cdot f_{N}(N_N)=\lambda\prod_{i=2}^N M_i, \quad \lambda <M_1
\end{equation}
The equation $f_1(N_1)=\lambda$ has just two real solutions, let us call them $N_1^-$ and $N_1^+$, since $f_1(N_1)$ increases from zero to $M_1$ and then decreases to zero at infinity. Clearly one has $N_1^-<w_1/A$ and $N_1^+>w_1/A$. It follows that if $N_1=N_1^-$ or $N_1=N_1^+$ equation (\ref{prodeq1}) has just one real solution  (the other $N_i$ being given by $N_i=w_i/A$, $i=2,...,n$). When $N_1<N_1^-$ or $N_1>N_1^+$ equation (\ref{prodeq1}) has no solution, since one has 
\begin{equation}
f_2(N_2)\cdot \ldots \cdot f_{N}(N_N)=\frac{\lambda}{f_1(N_1)}\prod_{i=2}^N M_i > \prod_{i=2}^N M_i.
\end{equation}
Finally it remains the case $N_1\in (N_1^-, N_1^+)$. In this case, having fixed the value of $N_1$, we can repeat the above reasoning on $N_2$, since now we have
\begin{equation}\label{prodeq1} 
f_2(N_2)\cdot \ldots \cdot f_{N}(N_N)=\frac{\lambda}{f_1(N_1)}M_2\prod_{i=3}^N M_i =\hat{\lambda}\prod_{i=3}^N M_i, \quad \hat{\lambda} <M_2
\end{equation}
and there will be two real solutions to the equation $f_2(N_2)=\hat{\lambda}$, $N_2^-<w_2/A$ and $N_2^+>w_2/A$. By repeating the same until the last $N$, we see that equation (\ref{prodeq1}), for $N_i>0, i=1,...,N$, represents a compact closed surface, isomorphic to the $N$-sphere. 
We just give two examples. Firstly, we choose again $N=3$ so that we can also plot the complete surface. We take
\begin{equation}\label{ex}
\frac{N_1^{2}}{e^{2 N_1}}\frac{N_2^{3}}{e^{2 N_2}}\frac{N_3^{4}}{e^{2 N_3}}=0.001.
\end{equation}
The maximum of the function $\frac{N_1^{2}}{e^{2 N_1}}\frac{N_2^{3}}{e^{2 N_2}}\frac{N_3^{4}}{e^{2 N_3}}$ equals to $54 e^{-9} \sim 0.0067$ so equation (\ref{ex}) possesses solutions. The corresponding surface is plotted in figure (\ref{fig4}). The next example is about $N=4$. We take:
\begin{equation}\label{ex1}
\frac{N_1^{2}}{e^{2 N_1}}\frac{N_2^{3}}{e^{2 N_2}}\frac{N_3^{4}}{e^{2 N_3}}\frac{N_4^{3}}{e^{2 N_4}}=0.0005.
\end{equation}
Now, the maximum of the function $\frac{N_1^{2}}{e^{2 N_1}}\frac{N_2^{3}}{e^{2 N_2}}\frac{N_3^{4}}{e^{2 N_3}}\frac{N_4^{3}}{e^{2 N_4}}$ equals to $729/4 e^{-12} \sim 0.00112$. In order to get a plot in three dimension, we fix the value of $N_4$ to be equal to 2 (since the value of $N_4^3/e^{2N_4}$ must be greater than $0.0005 e^{9}/54 \sim 0.075$). The corresponding closed surface, projected in the three dimensional space, is given in figure (\ref{fig4}). 
Thanks to our approach we have been able to construct a new  (we did not find anything analogous in the literature)	$N-1$ parameter family of integrals of motion, written in terms of the original dynamical variables, and moreover we  give a proof of the fact that one can extract from it $N-1$ first integrals in involution.
\section{Concluding remarks.}
A general comment is needed here, on the comparison between our results and those displayed  on pages
245-250  of \cite{Volterra2}, where the three basic properties of fluctuations in conservative associations are stated and explained.
There, first of all Volterra distinguishes between  even and odd number of species. In the odd case in fact the coefficient matrix $A$, being  skew symmetric, has to be singular: Volterra concludes that  for nonzero growth coefficients equilibrium states will be impossible, and the number of individuals in some species will grow indefinitely or go to zero, and on the long run, only an even number of species will survive. Curiously enough, Volterra did not take into account the fact that   complete integrability could change, even drastically, the above scenario. In fact, in section 3 we have shown  that even for an odd number of species with nonzero growth coefficients, in spite of the singular nature of the matrix $A$, there are bounded trajectories and  periodic orbits, and we conjecture that such behavior is actually valid for any N, even or odd, just because of integrability.
In Appendix A we will come back to the approaches followed in \cite{Telmo Jorge Lucas Peixe, LO, Baigent} showing that complete integrability holds true for a different Poisson structure, and relies  just to the assumed specific form of the matrix $A$.
%Here, we want once more to stress that we have shown that bounded, and indeed periodic, orbits do exist even  for an odd number of species: in fact they exist, thanks to complete integrability, whatever be the number of species.
% It is somehow surprising  that the results we have derived do not really rely upon the Hamiltonian formulation  of integrable systems; as a matter of fact, even the results we have got %for the two species case could have been derived without resorting to the Hamiltonian formulation. Moreover,  ubiquitous features in the mathematical picture of integrable system, %such as Lax representation, do not appear in our description.
%A further question, more important in view of the possible applications, is related to the construction of a richer model, still \tcr{conservative and thus amenable to a Hamiltonian description}, but involving a larger number of conjugated pairs of variables, all of them being significant in the economic-ecological approach. Work is progress in \tcr{this  direction}.
\medskip
\section*{Appendix A.}
In this Appendix we investigate the following issues:
\begin{enumerate}
\item
For completeness, we identify the analytic expression of the Hamitonian of the $N$-species Volterra system associated to the degenerate  Poisson bracket (see also \cite{LO}):
%As a matter of fact, we will show that they (or, as usual, a simple linear combination of them) will remain in involution even we employ a different Poisson bracket, for instance the one introduced in (\cite {LO}), , namely: 
\begin{equation}
\{f,g\} = 	\sum_{jk} N_j N_k A_{j,k}\frac{\partial f}{\partial N_j}\frac{\partial g}{\partial N_k} \label{fernandez}
\end{equation}
%which, through the change of variables $N_j:=\exp(y_j)$ simplifies to:
%\begin{equation}
%\{f,g\} = \sum_{jk}A_{j,k}\frac{\partial f}{\partial y_j}\frac{\partial g}{\partial y_k }\label{fernandezmod}
%\end{equation}
\item
We will  show that out of the family of first  integrals
\begin{equation}
e^{-\sum_{k=1}^n B_k\epsilon_{k}w_k\left(\sum_i N_i(t)\right)}\prod_{i=1}^n N_i(t)^{w_i}=I_{1,..,n}.\label{eqnewit}
\end{equation}
\noindent
one can extract $N-1$ independent integrals of motion, in involution with respect to the Poisson bracket (\ref{fernandez}).
\end{enumerate}
\begin{enumerate}
\item
For easiness of reading, we remind the explicit form of the $N$-species Volterra equations:% in terms of the exponential variables $y_j\equiv \exp(N_j)$:
\noindent
%\begin{equation}
%\dot y_j =\epsilon_j + \sum_j \epsilon_j\epsilon_k(B_j-B_k)\exp(y_k)\label{exp}
%\end{equation}
\begin{equation}
\dot{N}_r = \epsilon_r N_r + \sum_{s=1}^N A_{rs}N_rN_s ~~(r=1,\cdots,N)\label{exp}
\end{equation}
\noindent
and look for a Hamiltonian function  ${\mathcal K} $ such that (\ref{exp}) can be cast in the form:
%$$\dot y_j = \sum_k \epsilon_j\epsilon_k(B_j-B_k){\frac{\partial {\mathcal K}_1}{\partial y_k}}.$$
%i.e.
$$\dot{N}_r =\{N_r,\mathcal K\}$$
\noindent
where the Poisson bracket is given by (\ref{fernandez}).
\noindent
Notice that we cannot expect this Hamiltonian to be uniquely defined, as  Ker$(A)$ is non-empty, and thus  is defined up to Casimir functions. 
\noindent
%{\bf Proposition 1}.  
The family of Hamiltonians ${\mathcal K}$ for the system (\ref{exp}) is given by:
\begin{equation}
\mathcal K = \sum_{k=1} N_k - N_k^{0} \log(N_k)\label{hamK}
\end{equation}
\noindent
where  the constant coefficients $N_k^{0}$ belong to the equilibrium configuration and thus satisfy $\epsilon_k + \sum_{s=1}^N A_{ks}N_s ^{0}= 0$. In the three-species case, a possible parametrization is given by equations (\ref{KerA}).
The proof is by direct computation.
\item
Since we are interested in the involutivity, we can take a function of the integrals (\ref{eqnewit}). Let us take the logarithm, by defining: 
\begin{equation}
J (w_1\, \cdots, w_N) =\sum_{k=1}^N \epsilon_kw_k D_k 
\end{equation}
\noindent
where the coefficients $D_k$ read:
$$D_k = (1/\epsilon_k)\log(N_k) -B_k\sum_jN_j.$$
\noindent
We are obviously  assuming that the matrix $A$ is not just skew-symmetric, but of the form 
$A_{r,s}=\epsilon_r\epsilon_s(B_r-B_s)$. A troublesome though straightforward calculation yields
$$\{D_k,D_j\} = \left(B_k-B_j\right)\left(1-\sum_{i}\epsilon_iB_i N_i\right)$$
implying once again that the Poisson brackets $\{D_j-D_r, D_k-D_r\}$ ($r$ fixed, $j,k$ running from $1$ to $N$) vanish, then providing $N-1$ integrals of motion in involution. 
\end{enumerate}
\noindent
So, we have been able to write the Hamiltonian and the integrals of motion in terms of the biological variables. We notice that Volterra was the only one who derived the special, and singular form of the matrix $A$ that ensures complete integrability.
%%%%%%%%%%%%%%%%%%%%%%%%%%%%%%%%%%%%%%%%%%%%%%
\subsection*{Acknowledgements}
F.Z. wishes to acknowledge the support of Universit\`a degli Studi di Brescia; INFN, Gr. IV - Mathematical Methods in NonLinear Physics and ISNMP - International Society of Nonlinear Mathematical Physics. O.R. and F.Z. wish to acknowledge the support of GNFM-INdAM.
%%%%%%%%%%%%%%%%%%%%%%%%%%%%%
\label{lastpage}
\end{document}